\documentclass{aa}
\usepackage{natbib}
\bibpunct{(}{)}{;}{a}{}{,}
\usepackage{graphicx}
\usepackage{txfonts}
\usepackage{longtable,lscape}
\begin{document}
   \title{Candidate planetary nebulae in the IPHAS photometric catalogue}

   \author{K. Viironen\inst{1} \and
          R. Greimel\inst{2} \and
          R. L. M. Corradi\inst{1} \and
          A. Mampaso\inst{1} \and          
	  M. Rodr{\'{i}}guez\inst{3} \and
	  L. Sabin\inst{4} \and
          G. Delgado-Inglada\inst{3}
	  J. E. Drew\inst{5} \and
          C. Giammanco\inst{1} \and          
          E. A. Gonz\'alez-Solares\inst{6} \and
          M. J. Irwin\inst{6} \and
          B. Miszalski\inst{7,8} \and
          Q. A. Parker\inst{7}\and
          E. R. Rodr{\'{i}}guez-Flores\inst{9,1} \and
	  A. Zijlstra\inst{4}
          }

   \offprints{K. Viironen}

   \institute{Instituto de Astrof{\'{i}}sica de Canarias (IAC), C/V{\'{i}}a L\' actea s/n,
              38200 La Laguna, Tenerife, Spain\\
              \email{kerttu@iac.es}
         \and
	 Institut f\"ur Physik, Karl-Franzens Universit\"at Graz, Universit\"atsplatz 5,8010 Graz, Austria\\
	 \and
	 Instituto Nacional de Astrof{\'{i}}sica, \'Optica y Electr\'onica INAOE, Apdo Postal 51 y 216, 72000 Puebla, Pue., Mexico\\
         \and
	 Jodrell Bank Centre for Astrophysics, Alan Turing Building, University of Manchester, Manchester, M13 9PL, UK\\
	 \and
         Centre for Astrophysics Research, University of Hertfordshire, College Lane, Hatfield AL10 9AB\\
	 \and
         Institute of Astronomy, Madingley Road, Cambridge CB3 0HA\\
         \and
	 Department of Physics, Macquarie University, Sydney, Australia \\
         \and
         Observatoire Astronomique, Universit\'e de Strasbourg, 67000, Strasbourg, France \\
         \and
         Instituto de Geof\'{i}sica y Astronom\'{i}a, Calle 212, N. 2906, CP 11600, La Habana, Cuba
     }

   \date{}


  \abstract
    {We have carried out a semi-automated search for planetary nebulae (PNe) in the INT Photometric H-Alpha Survey (IPHAS) catalogue. We present the PN search and the list of selected candidates. We cross correlate the selected candidates with a number of existing infrared galactic surveys in order to gain further insight into the nature of the candidates. Spectroscopy of a subset of objects is used to estimate the number of PNe present in the entire candidate list.}
    {The overall aim of the IPHAS PN project is to carry out a deep census of PNe in the northern Galactic plane, an area where PN detections are clearly lacking.}
    {The PN search is carried out on the IPHAS photometric catalogues. The candidate selection is based on the IPHAS and 2MASS/UKIDSS colours of the objects and the final candidate selection is made visually.}
    {From the original list of $\sim$ 600 million IPHAS detections we have selected a total of 1005 objects. Of these, 224 are known objects, leaving us with 781 PN candidates. Based on the initial follow-up spectroscopy, we expect the list to include very young and proto-PNe in addition to genuine, normal PNe ($\sim$16\%) and emission line objects other than PNe. We present additional criteria to select the most probable PN candidates from our candidate list.}
    {}


   \maketitle
%

\section{Introduction}

Planetary nebulae (PNe) are the evolutionary end products of most low and intermediate mass stars (approximately 1 to 8 solar masses). About 2700 PNe have been detected so far in the Galaxy, including the objects listed by \cite{acker94,acker96} and the new PNe found in the Southern Hemisphere by the AAO-UKST H$\alpha$ Survey \citep{parker05,miszalski08}. However, the expected Galactic PN population is much larger, with estimates varying from $28000\pm5000$ based on observations \citep{frew06} to $46000\pm13000$ PNe (of size $<0.9$ pc) derived from stellar population synthesis models \citep{moe06}. In particular, there is a clear lack of PN detections in the centre of the Galactic Plane \citep[see for example Fig. 6c in][]{miszalski08} due to difficulties in observing nebular objects in areas of high extinction and where confusion with other types of nebulae (e.g. H\,{\sc ii} regions) is significant.

IPHAS is the northern counterpart of the AAO-UKST H$\alpha$ Survey which mapped the Southern Galactic plane ($\mid b \mid$ $\leq$ 10 degrees) and discovered $\sim$ 1200 new PNe \citep{parker06,miszalski08}. Both of the surveys are of similar depth (2-5 Rayleighs) for diffuse, extended emission. However, the resolution and photometric quality of IPHAS is better, making it more suited to search for compact PNe.

IPHAS mapped 1800 degrees$^2$ of the Northern Galactic Plane (a band between $b$= --5 to +5 degrees) in three filters using the INT Wide Field Camera at the Observatorio del Roque de los Muchachos (La Palma, Spain). A narrow-band H$\alpha$ filter (central wavelength and width: 6568~\AA/95~\AA) and two Sloan filters ($r^\prime$ and $i^\prime$) were used for matched 120, 30, and 10 s exposures, respectively. For point sources the survey covers the magnitude range from $r^\prime \sim$ 13 to 22 mag (5$\sigma$ detection limit, magnitudes are in the Vega system) with the median magnitude limit being $r^\prime$=21 mag. Each IPHAS field is observed twice at two closely overlapped pointings. More information about the survey can be found in \cite{drew05}. 

To search for new PNe in IPHAS, we use two search methods. A semi-automated method was used based on IPHAS photometry to find mainly small angular diameter PNe (typically $\le 5\arcsec$), and visual inspection of IPHAS mosaics to find mainly extended PNe (typically $\ge 5\arcsec$). The two methods overlap in the size of objects found, leading to a complete search for PNe with diameters from subarcseconds to many arcminutes. Here we present the search method and the results of the semi-automated PN search. The results of the visual PN search will be published in the near future \citep[][S09 from now on]{corradi09,sabin09}.

A search for barely resolved or point-like PNe based on a catalogue of 4853 H$\alpha$ emitting stars \citep{witham08} was published in \citet{viironen09} (V09 from now on). Here we carry out the search in the complete IPHAS photometric catalogue, covering almost the whole Northern Plane including also non-stellar objects and objects without detection in the $i^{\prime}$ band. We aim at picking up all the possible PNe included in the photometric catalogue but due to the source detection technique used in the catalogue generation \citep{gonzalez-solares08}, we are more likely to discover compact PNe. In addition to the overall aim of the IPHAS PN project, gaining better understanding of the total Galactic PN population, we are especially interested in the compact PNe as they are probably either very young or distant objects. Thanks to the good spatial resolution of IPHAS, we are able to resolve subarcsecond objects which are ideal candidates of young or proto-PNe. These,
  in turn, give us valuable information about the poorly known late stages of stellar evolution of the PN progenitor stars. 

The discovery of even a few new very young PNe (vyPNe) would be important
as only a few of them are known and not many are expected in the
Galaxy. For example, assuming that the PN lifetime is 20000 years and
the estimated Galactic population 30000, we can expect that the
Galactic population of PNe younger than 1000 years is 1500. Of the
expected PNe, only $\sim$2700 are detected and of these about 200 are
in the IPHAS area. If the constraints of the typical searches for
large PNe and for our search of vyPNe were the same, we could expect to
find only $\sim$ 10 new vyPNe in the IPHAS area. But as our search is
much deeper and has better spatial resolution than previous searches, 
we expect to find a larger number of vyPNe.


\begin{figure*}[!t]
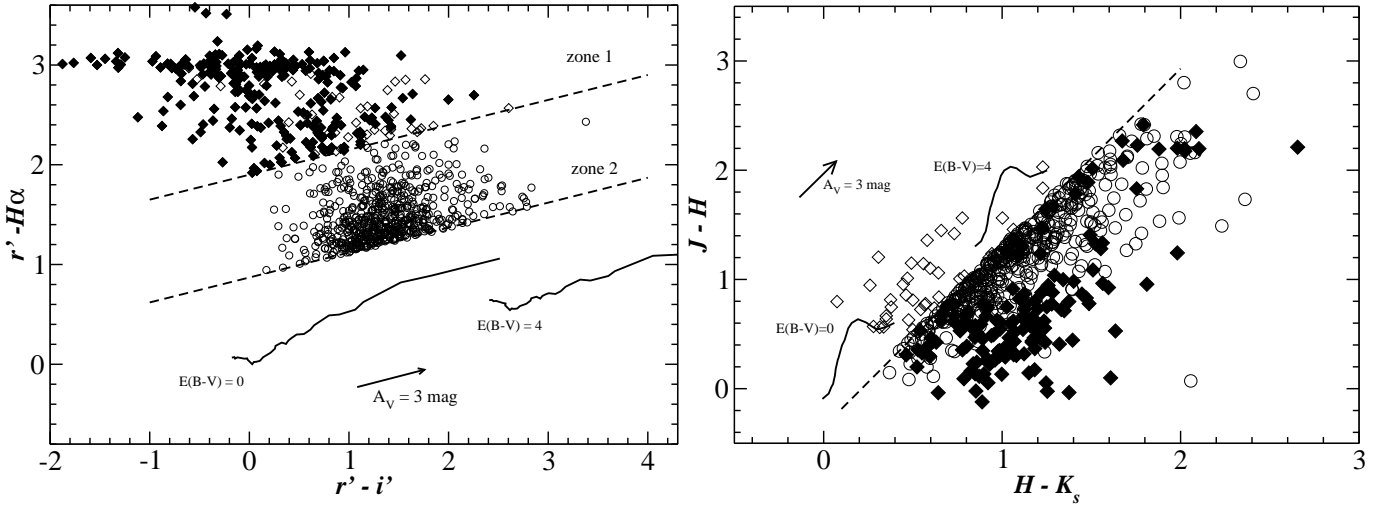

   \centering
   \includegraphics[width=\columnwidth]{Fig1a.eps}
   \includegraphics[width=\columnwidth]{Fig1b.eps}
      \caption{Two-colour diagrams -- {\it Left} IPHAS colours, {\it Right} 2MASS colours. Shown are the locations of the IPHAS selected PN candidates (diamonds). The ones located also below the cut line in 2MASS are shown as filled diamonds. IPHAS+2MASS selected PN candidates are shown as circles. Main sequence tracks at different reddenings are shown as solid lines and the colour cuts applied in the candidate selection as dashed lines (see text).}
\label{fig:ip2mdiagram}
\end{figure*}

\section{Search method}\label{s2}

The source of the semi-automated PN search is the IPHAS object catalogue, including all the IPHAS observations until January 2008 (inclusive), thus covering 93\% of the total IPHAS area. The initial data release has been published and a detailed description of the data reduction and products is provided in \cite{gonzalez-solares08}. 

Our PN candidate search consists of cleaning the catalogue data from possible false detections and of choosing the PN candidates due to their location in the IPHAS and Two Micron All Sky Survey \citep[2MASS, ][]{skrutskie06} colour-colour diagrams. 

\subsection{The catalogue search}

The IPHAS catalogue includes about 600 million detections of Northern plane objects, most of them detected two or more times. We require detections at least in the H$\alpha$ and $r^\prime$ filters. We can assume that a PN is visible in the continuum $r^\prime$ filter due to presence of the H$\alpha$ line while a detection in the $i^\prime$ filter is not assured. We also require a matching distance between the detections in different filters of $<1.4\arcsec$ for nebular objects and $<1\arcsec$ for stellar objects. These limits were defined empirically from IPHAS detections of known PNe. Based on the known IPHAS colours of PNe (see V09) we start our candidate selection by making a generous colour cut: $r^\prime - \mbox{H}\alpha > 0.25(r^\prime - i^\prime) + 0.55$. This cut line was chosen to include all the known PNe in the IPHAS two-colour diagram while cutting out the main sequence of stars. The line was aligned along the approximate reddening vector \citep[see][]{corradi08}. To remove possible artifacts, we discard detections at the borders of the CCDs and in areas of bad pixels. In the IPHAS catalogue, all detections are classified as either Saturated, Star, Probable star, Probable extended, Extended or Noise \citep[see Table 7 in][]{gonzalez-solares08}.
We remove detections classified as Noise or Saturated, except those that are saturated only in H$\alpha$ in order not to eliminate possible very bright H$\alpha$ emitters.
After these initial selection steps, we reduced the number of possible candidates to $\sim14$ million.

Most of the possible candidates after the initial selection clearly have to be false detections. 
Studying the initial selection, we found the following problems:
in cloudy nights, thicker cloud cover in the r$^\prime$ than in the
H$\alpha$ image can cause apparent H$\alpha$ emitters;
areas of diffuse H$\alpha$ emission contain detections of arbitrary nebular objects;
in some cases a satellite crossing the H$\alpha$-image is matched with a star at the same position in $r^\prime$ and $i^\prime$ causing it to be detected as an H$\alpha$ emitter.
To remove these kinds of problems, we require that the objects must be detected at least twice and at least one of these two observations must be in a field that fulfils the IPHAS quality criteria, in line with the PhotoObjBest requirements in \citet{gonzalez-solares08}. About 99\% of the IPHAS objects are observed at least twice, which allows us to apply this selection. For bad quality fields, their observation will be repeated later so that the objects possibly lost in this step will eventually be recovered. After these selection steps we are left with $\sim1.3$ million detections.

So far we have been working with individual IPHAS detections. As a next step, we average the magnitudes of the detections for each object. For this we use only the magnitudes in the fields fulfilling the IPHAS quality criteria. In the case that an object has an $i^\prime$-measurement in some of the observations but not in others, only the good quality images with $i^\prime$ detection are averaged. We also make a cut in H$\alpha$ magnitude, m(H$\alpha$)$< 19$. Our motivation for this magnitude cut is that the limiting magnitude of IPHAS is m($r^\prime$) $\simeq$ 20-22 \citep{gonzalez-solares08}. The theoretical maximum IPHAS $r^\prime$ - H$\alpha$ colour of a pure H$\alpha$ emitter (assuming that all the flux in the $r^\prime$ filter comes from the H$\alpha$ line) is 3.1 \citep{drew05}.
Therefore to avoid being biased towards objects with smaller H$\alpha$ excess at faint magnitudes, the magnitude cut is a reasonable choice.
This also reduces problems caused by rising photometric errors at fainter magnitudes. After the averaging and the magnitude cut, we are left with $\sim165000$ objects.

At this stage, the most frequent remaining false detections are close
optical binaries. Often these objects are not resolved in H$\alpha$
because of the slightly broader PSF in this filter due to the longer exposure time. Hence the binary is measured as an extended object in the H$\alpha$ image while it is resolved in $r^\prime$ and $i^\prime$. The two stars in $r^\prime$ and $i^\prime$ are then matched with the same 'nebular' H$\alpha$ object. We removed these false detections by eliminating all objects that have two $r^\prime$ detections within 2$\arcsec$ for the same H$\alpha$ detection. The radius of 2$\arcsec$ was defined empirically. After removing this kind of false emitter, we are left with 88000 objects. For these objects, the next selection step is based on their IPHAS and 2MASS colours.

\subsection{Colour criteria}

After having cleaned the candidates using the above steps,
near-infrared counterparts of the remaining IPHAS objects were
sought. A matching radius of 1.4$\arcsec$ was used when searching for
matches in the 2MASS Extended Source Catalogue, while for the Point
Source Catalogue a 1$\arcsec$ search radius was used. Note that the
external astrometric precision of IPHAS with respect to 2MASS is
generally better than 0.1 arcsec \citep{gonzalez-solares08}. If the
object has both an extended and a point source counterpart, the latter
was selected. In the case that no 2MASS counterpart was found,
UKIDSS\footnote{The UKIDSS project is defined in
  \citet{lawrence07}. UKIDSS uses the UKIRT Wide Field Camera
  \citep[WFCAM,][]{casali07} and a photometric system described in
  \citet{hewett06}. The pipeline processing and science archive are
  described in \citet{irwin09} and \citet{hambly08}. We used data from the 4th data release.} data was used in the
colour selection instead.
Having IPHAS and 2MASS/UKIDSS information in hand, the PN candidates
were selected based on their locations in the IPHAS ($r^\prime$ -
H$\alpha$) vs. ($r^\prime$ - $i^\prime$) and 2MASS ($J - H$) vs. ($H -
K_S$) diagrams. No colour correction was applied to UKIDSS data as the
corrections relative to the 2MASS photometric system are small
\citep{hodgkin09}. The colour selection is based on the location of
known PNe and other classes of emission line objects and normal stars
in these diagrams. The diagrams were discussed in \cite{corradi08} and
V09 and the candidate selection method is the same here as in V09. The zone 1 and 2 (see Fig.~\ref{fig:ip2mdiagram}) introduced in V09 are defined by the following equations a) and b), respectively: 

\begin{displaymath}
\mbox{a)}\;\; r^\prime-\mbox{H}\alpha > 0.25(r^\prime - i^\prime) + 1.9 \,\,\,\mbox{or}
\end{displaymath}

\begin{displaymath}
\phantom{a)}\;\; r^\prime-\mbox{H}\alpha > 1.9 \quad\mbox{(for objects without $i^\prime$ magnitude)}
\end{displaymath}

\begin{displaymath}
\mbox{b)}\;\; 0.25(r^\prime - i^\prime) + 0.87 < r^\prime-\mbox{H}\alpha < 0.25(r^\prime - i^\prime) + 1.9 \,\,\,\mbox{or}
\end{displaymath}

\begin{displaymath}
\phantom{b)}\;\; 0.87 < r^\prime-\mbox{H}\alpha < 1.9 \quad\mbox{(for objects without $i^\prime$ magnitude)}
\end{displaymath}

\begin{displaymath}
\phantom{b)}\;\;\mbox{and} 
\end{displaymath}

\begin{displaymath}
\phantom{b)}\;\; J-H<1.64(H-K_s)-0.35.
\end{displaymath}

The objects that fulfil the IPHAS criterion a) and the 2MASS criterion b) have the most probable colours to be genuine PNe. To be more complete in these highest probability zones, we also selected the objects having only a single IPHAS detection for visual inspection if they fulfil these colour criteria.

After the colour selection we are left with 4740 potential PN candidates.

\subsection{Visual inspection}

As a final selection step, all 4740 objects were checked visually in their IPHAS preview images. With a few exceptions, the remaining visually distinguishable mimics are detections in areas of large H$\alpha$ emitting nebulae such as H\,{\sc ii} regions or super-nova remnants.
In some cases, a faint star projected on an extended nebulosity can
cause a false H$\alpha$ emitter detection because of poor subtraction
of the highly variable background. If this is clearly the case, these objects are removed, otherwise the object was retained. In general, if the non-PN nature was clear from the morphology, the object was discarded from the candidate list in this visual study. If not, the object was included in the candidate list and the final decision on its nature left for future spectroscopic studies. Very extended objects are discarded, as the morphology is not well seen in the preview images. These objects will be recovered in the extended nebula search (S09) of IPHAS image mosaics. Generally this size limit is $\sim$1$\arcmin$ but a few more extended nebulae with bright cores are included in our candidate list.

\begin{figure}[!t]
   \centering
   \includegraphics[width=0.33\linewidth,angle=-90]{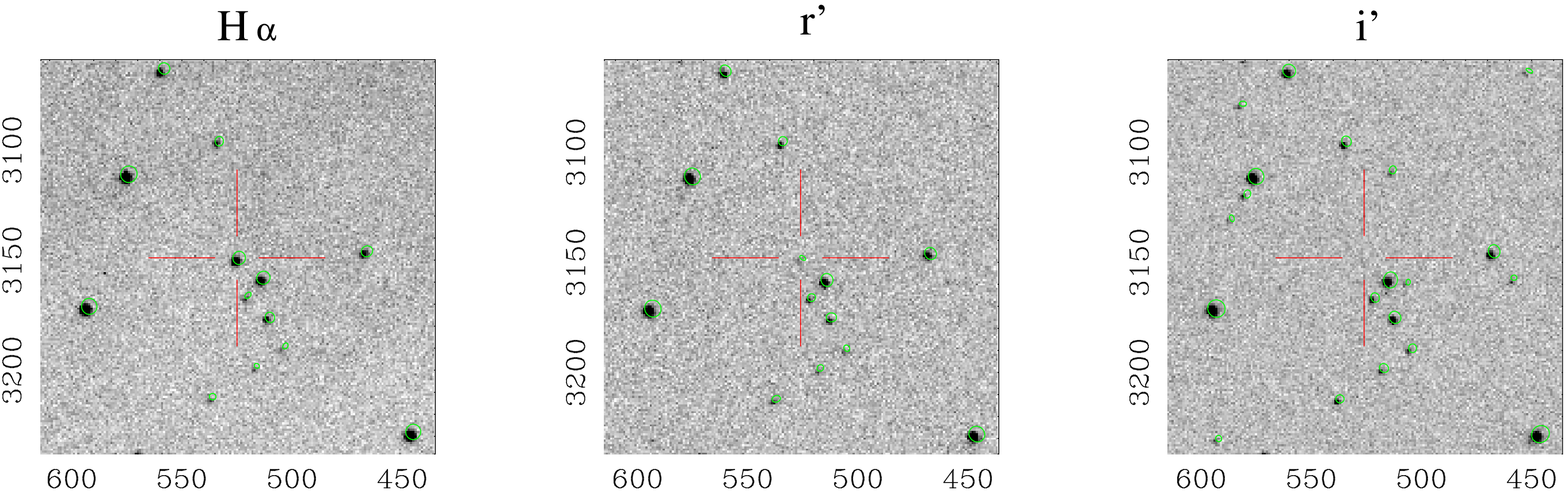}
   \includegraphics[width=0.33\linewidth,angle=-90]{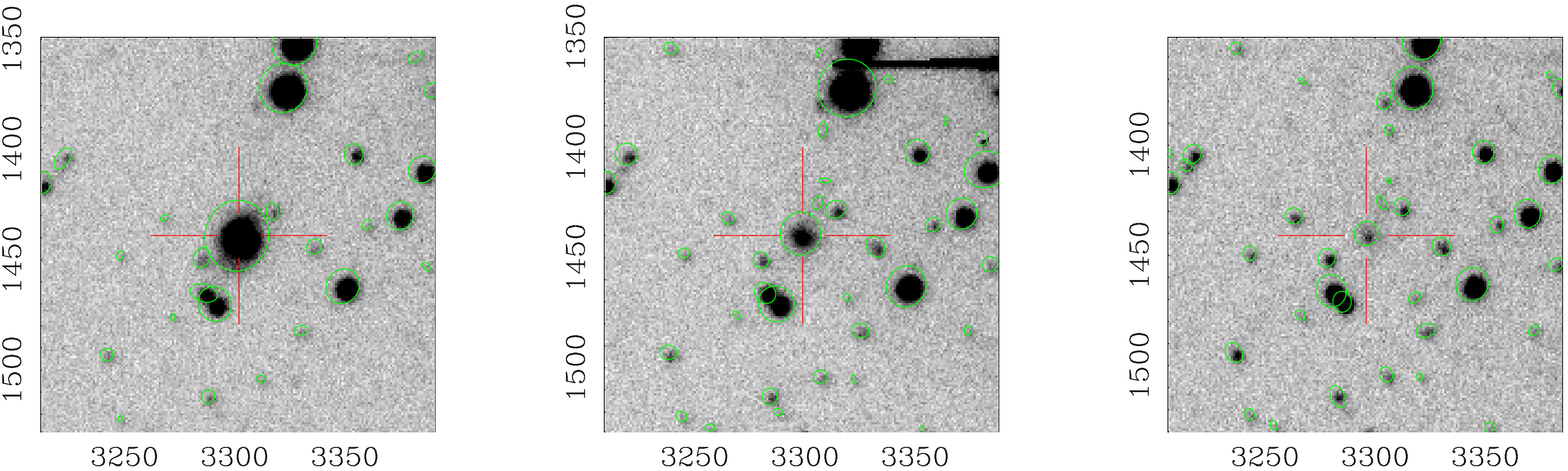}
   \includegraphics[width=0.33\linewidth,angle=-90]{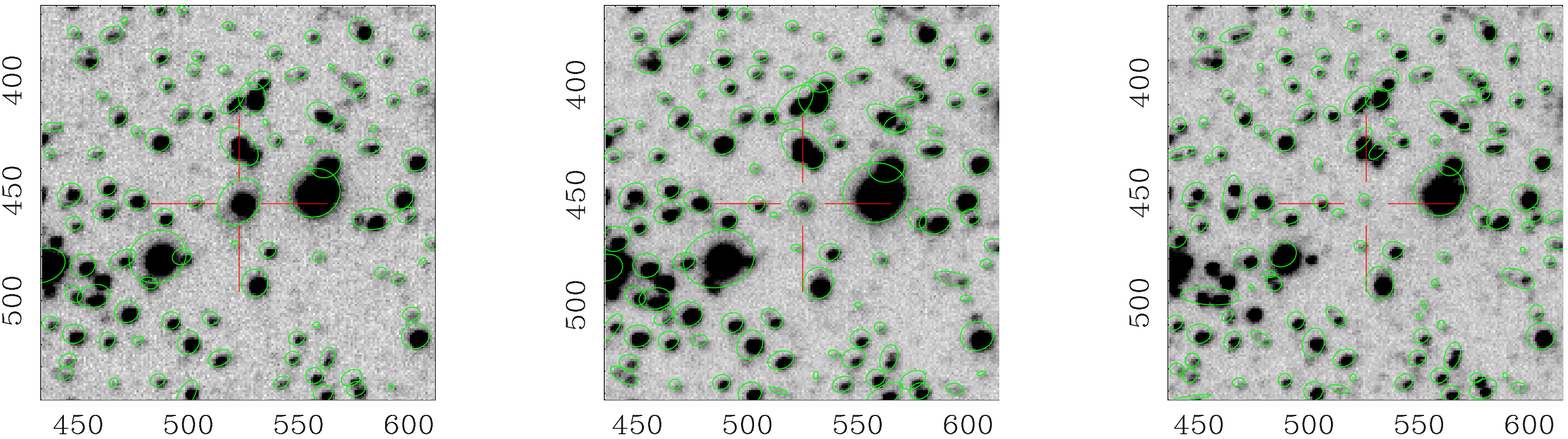}
   \includegraphics[width=0.33\linewidth,angle=-90]{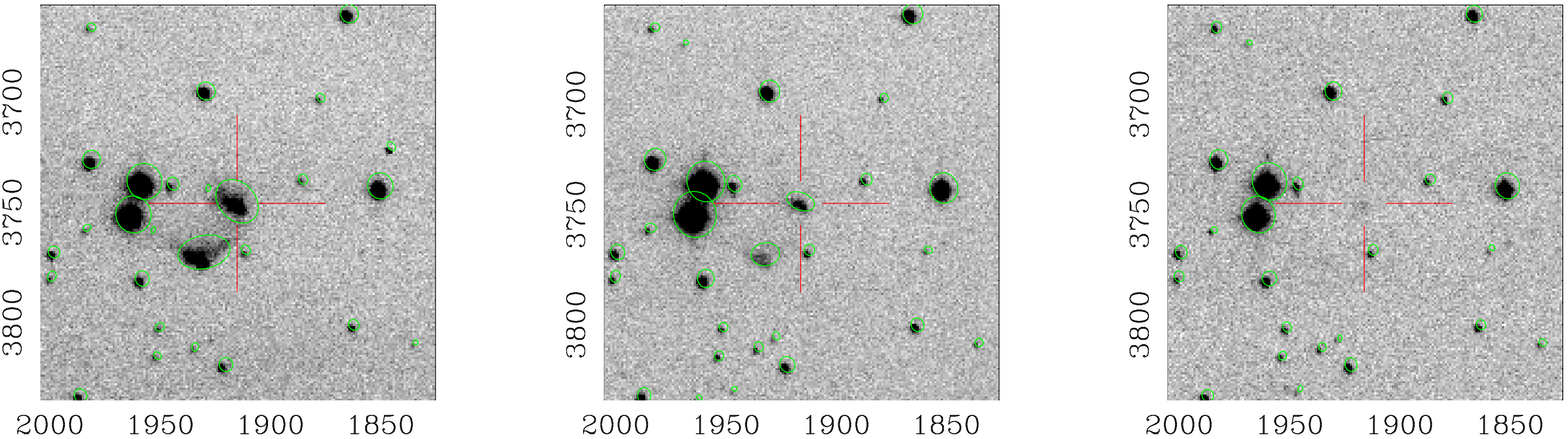}
      \caption{H$\alpha$, $r^\prime$ and $i^\prime$ IPHAS preview images, from left to right, of a few PN candidates. The image sizes are $1\arcmin \times 1\arcmin$, north is up and east towards left. The IPHAS detections are shown as circles. The three images in the bottom row illustrate the selection of a condensation in a new extended PN candidate.}
\label{fig:gooddetections}
\end{figure}

\subsection{Results}

The final result of our PN search is a list of 1005 objects. These are listed in Table 1 which is available only electronically. We searched first for possible SIMBAD entries for our candidates and found 311 matches. Of these, 168 have a primary object type of PN or possible PN. Other relatively frequent objects worth mentioning are 34 emission objects or emission line stars, of which three have as a secondary object type PN and according to the literature two of these are indeed PNe, 18 H\,{\sc ii} regions, 19 Herbig-Haro (HH) objects, 13 nebulae of unknown nature, 11 galaxies and 11 young stellar objects (YSOs), one being classified as candidate only. Our selection also picked out six known novae. We have added the SIMBAD information to our final candidate table. We also checked all references in the literature for the 311 objects classified in SIMBAD. In 224 cases we considered the object classification well established, and we did not study these objects any further.

  The 87 objects for which we considered that the object nature is not well known are further studied together with the 694 candidates without any SIMBAD classification. Thus, we are left with a total of 781 PN candidates. Of the 83 candidates from the list of \citet{witham08} which were already published in V09 (as marked in Table 1), all except 10 objects are selected here as well. The missing objects were not selected because the \citet{witham08} selection was made before a recent change in the H$\alpha$ filter calibration had been implemented and the objects now fall outside our colour cuts or in the fields not fulfilling the IPHAS quality criteria.

In addition to compact candidates, we also found many extended objects due to catalogued condensations in larger nebulae. In these cases, in the final candidate list, the brightest (in H$\alpha$) catalogued condensation and the corresponding magnitudes are listed. In addition, the coordinates of the centre and the largest extension of the nebula are given. A detailed study of the extended nebulae in IPHAS will be published later (S09). Examples of new IPHAS PN candidates are shown in Fig.~\ref{fig:gooddetections}.

\section{Completeness and reliability of the catalogue}\label{analysis}

The catalogue we base our study on covers 93\% of the IPHAS area. The PN search in the remaining 7\% will be carried out when the data are available. Our aim is to discover as many new Galactic PNe as possible. However, as the search presented here is carried out in the IPHAS photometric catalogue, we can aim to be complete only for compact PNe ($\leq 5 \arcsec$). Of these objects, we expect to lose 4\% in our candidate selection process: 1\% due to the requirement of a minimum of two detections and 3\% due to the applied colour selection.

We tested this in practice by checking if the 63 known compact ($\leq 5 \arcsec$) PNe \citep{acker94,parker06,miszalski08} in the IPHAS area and inside the IPHAS brightness limits were selected by our algorithm. Six of these were not picked up: 4 because of problems in the IPHAS catalogue generation which sometimes splits a single extended object into several objects, and 2 due to our selection algorithm, in line with our prediction above.

We have carried out our search aiming to be as complete as possible
rather than to have a clean selection of PNe. Of the possible mimics, we expect that the most common are T-Tauri stars because they enter our colour selection (see V09 and
  Section~\ref{sec:ip2m}) and they can be expected to be numerous
  objects. Possible mimics are also symbiotic stars because these
  objects enter clearly into our colour selection (see V09) and their
  predicted number is large \citep[from $3\times10^3$ to $4\times10^5$
  Galactic systems, see][and references therein]{corradi08}. Further potential mimics are HH objects linked with
young stars, although these are discarded in the visual inspection if the morphology allows a clear separation from PNe. Also, massive young stellar objects (mYSOs) and H\,{\sc ii} regions are possible mimics. The latter were again usually discarded in the visual inspection, however compact H\,{\sc ii} regions especially might have remained in the final candidate list.

\begin{figure}
   \centering
   \includegraphics[width=\columnwidth,angle=0]{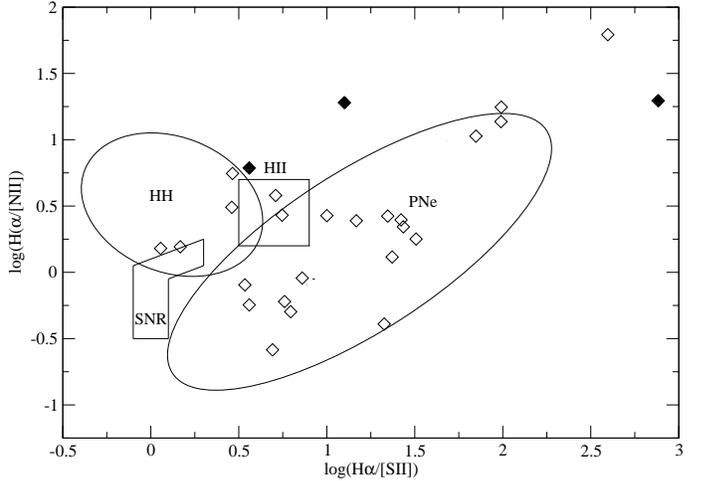}
      \caption{The S2N2 diagram of \citet{sabbadin77} showing the zones for SNRs and H\,{\sc ii} regions, the 0.85 probability ellipse for normal PNe \citep{riesgo06}, and the HH zone (V09). Over-plotted is our sample of 27 spectroscopically observed PN candidates (diamonds). Three new IPHAS vyPNe studied in V09 are shown as filled diamonds.}
\label{fig:S2N2}
\end{figure}
 
\section{The nature of the candidates}\label{sec:4}

\subsection{Follow-up spectroscopy}

In order to confirm the PN nature of our candidates, optical spectra
(as a minimum) are needed. For this purpose we have started a
follow-up spectroscopy campaign. So far we have spectra for 69 of the
PN candidates using the 2.1m telescope at the Observatorio Nacional de
San Pedro M\'artir (SPM), Mexico, Isaac Newton Telescope (INT),
William Herschel Telescope (WHT), and Nordic Optical Telescope (NOT),
La Palma, Spain, and the Australian National University 2.3-m
telescope at Siding Spring Observatory (SSO), Australia. These
  include the 19 spectra mentioned in V09. In this paper, the results
from the spectroscopic study are only used to assess the efficiency of our PN selection. Detailed study of the individual spectra will be published in subsequent papers. A few of these objects have been already published by \citet{parker06} or \cite{miszalski08} (as indicated in Table 1). However, as they were independently discovered from IPHAS, they will be included here in our follow-up spectroscopy statistics.

\subsubsection{New IPHAS PNe, symbiotic stars and emission line stars}

Of the 69 observed objects, 38 show a spectrum with nebular lines typical of normal PNe. We refer to these as IPHAS nebulae in the following discussion. Closer inspection of the 38 spectra led us to classify 4 of them as likely symbiotic stars \citep[for details about recognising symbiotic star spectra, see][]{corradi08}. We refer to these as IPHAS symbiotics. 

To analyse the nature of the remaining objects we placed them into the log(H$\alpha$/[N\,{\sc ii}]) vs. log(H$\alpha$/[S\,{\sc ii}]) (S2N2) diagnostic diagram (as discussed by V09), which is a revised version of the original S2N2 diagram by \citet{sabbadin77}. 27 of the remaining 34 objects show all the necessary emission lines in their spectrum, see Fig.~\ref{fig:S2N2}. Of these, 16 are located within the area of the 0.85 PN probability ellipse and outside the overlapping areas of other types of objects. We will refer to these 16 objects as IPHAS PNe.

It is important to emphasise that objects not included within the 85\% probability ellipse but showing typical PNe lines may also very well be genuine PNe. The same can be true for the 7 IPHAS nebulae not plotted in Fig.~\ref{fig:S2N2}. As a matter of fact, we have marked in Fig.~\ref{fig:S2N2} three new PNe discussed in V09 and they all are located outside the PN ellipse (one of the V09 new young PNe does not show [S\,{\sc ii}] lines in its spectrum and is thus not plotted here). The objects outside the ellipse are probably peculiar and a detailed study is needed in order to classify them.

The 31 clear non-PNe belong to other classes of H$\alpha$
  emitting objects such as YSOs. They are interesting newly discovered
  objects on their own and a study of their spectra is in progress
  \citep{Valentini} but in our search for new PNe
  they are set aside as mimics. We will refer to them as IPHAS emission line stars, representing the sample of undesired objects in our candidate selection.

\subsubsection{PN discovery rate}

The sample of objects with follow-up spectroscopy is small for robust
statistics but we estimate that about 16\% of the 781 candidates are
genuine, normal PNe. We derived this result with the following
reasoning. 28\% of the candidates are located in zone 1 in the
IPHAS two-colour diagram. There are 16 IPHAS PNe, 10 IPHAS emission
line stars and 3 IPHAS likely symbiotic stars in zone 1. From these
data we can deduce that 55\% of the zone 1 objects are PNe. On the
other hand, from the location of known PNe in the IPHAS two-colour
diagram (see V09), we know that the zone 1 objects represent 95\% of
the PN population -- the rest being in zone 2. This leads to the
quoted total PN match rate of $0.28\times0.55/0.95 = 0.16$.

The statistics can be biased by the fact that 38 out of 69 (55\%) of the spectra are of objects classified as stellar in IPHAS while 582 out of 781 new objects (75\%) are stellar according to our diameter study (see below). Also, the statistics are based, on the one hand, on properties of known PNe and on the other hand on properties of new normal PNe (i.e. the group of IPHAS PNe, new objects which are located inside the 85\% probability ellipse of PNe in the S2N2 diagram, Fig.~\ref{fig:S2N2}). In addition to this kind of normal PNe, we assume that our candidate list includes a fair number of young PNe, similar to those presented in V09. We note that three out of four of the new young PNe in V09 are located in zone 2 in the IPHAS two-colour diagram and in addition three of them are located outside the 85\% probability ellipse of PNe, while one of them does not show the [S\,{\sc ii}] doublet in emission in our spectrum.

However, we note that 90\% of our candidates are of diameter $\leq 5\arcsec$. Even a PN match rate of only 16\% means that our candidate list includes 112 compact PNe in the Northern Galactic plane. This would roughly triple the number of known compact PNe in the IPHAS area. Considering that we expect to discover a so far hidden population of vyPNe, like the four objects in V09, we assume this number to be even higher.

\begin{figure*}[!t]
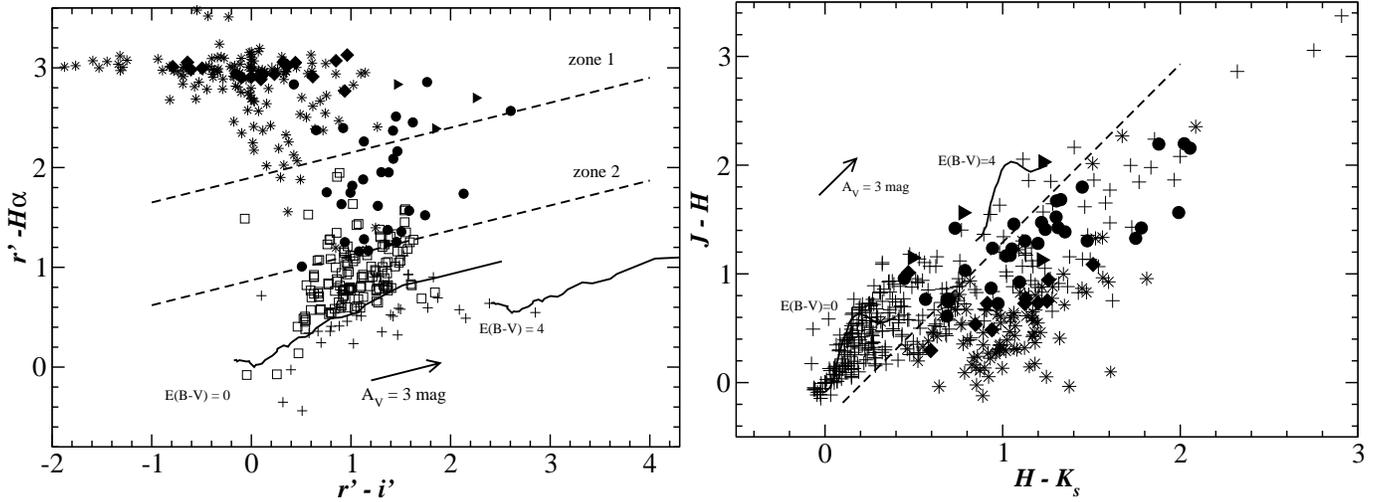

   \centering
   \includegraphics[width=\columnwidth,angle=0]{Fig4a.eps}
   \includegraphics[width=\columnwidth,angle=0]{Fig4b.eps}
      \caption{{\it Left} Location of the new IPHAS PNe (filled diamonds), new IPHAS emission line stars (filled circles), and new IPHAS symbiotic stars (filled triangles) in the IPHAS two-colour diagram. Plotted are also the locations of \citet{herbig88} T-Tauri stars (squares) and the post-AGB stars of \citet{szczerba07} (plus signs). The known PNe in our candidate list are shown as asterisks. Main sequence tracks are shown as solid lines and the colour cuts applied in the candidate selection as dashed lines (see text). {\it Right} 2MASS two-colour diagram. The same symbols and line-styles are used as in the left hand panel. T-Tauri stars are not plotted as their location was studied in \citet{corradi08}.}
\label{fig:ip2mdiagram2}
\end{figure*}

\subsection{Diagnostic diagrams}

In addition to the IPHAS and 2MASS two-colour diagrams, we studied the location of our candidates in various diagnostic diagrams combining IRAS \citep[The Infrared Astronomical Satellite,][]{beichman88}, 2MASS and MSX \citep[Midcourse Space Experiment,][]{egan99} infrared data, in cases where they were detected.

In all of the diagrams we have only plotted those PN candidates which have the needed fluxes well measured. We have separately indicated the locations of known PNe included in our candidate list. We have also marked the locations of the IPHAS PNe, IPHAS emission line stars, and IPHAS symbiotic stars. As the sample of new IPHAS objects (for which we have follow-up spectroscopy) is small, we have plotted them even when only a limit of the flux is available. In these cases, arrows indicate the direction for the possible location of the source in the diagram. As we expect our major mimics to be T-Tauri stars and symbiotic stars, we have studied the location of these objects in those diagrams where this work has not previously been done by other authors. For this purpose we have searched the data for the symbiotic star sample of \citet{belczynski00} and the T-Tauri stars of \citet{herbig88}.

As mentioned above, in addition to normal PNe, we expect that our candidate list includes vyPNe or even proto-PNe, like the four objects presented in V09. To get an idea about the expected locations of these kinds of objects in the diagnostic diagrams, we have also plotted in them the 'very-likely post-AGB stars' of \citet{szczerba07}. 

\subsubsection{IPHAS and 2MASS}\label{sec:ip2m}

The IPHAS and 2MASS two-colour diagrams for the objects for which we have spectra are shown in Fig.~\ref{fig:ip2mdiagram2}. The error bars are not plotted but are given in Table 1. However, generally the errors are about the same size as the symbols used. The locations of the newly discovered objects are in line with the locations of the previously known objects (see V09). The location of T-Tauri stars in IPHAS was not shown in V09 and here we see that some T-Tauri stars enter our zone~2 in the IPHAS diagram. 

Only 42 post-AGB stars from \citet{szczerba07} have an IPHAS
  counterpart. These objects do not necessarily yet show H$\alpha$ in
  emission which is apparent in the IPHAS two-colour diagram. Below
  our selection line in 2MASS, the post-AGB objects are located towards higher $J-H$ and $H-K_s$ colours as compared to PNe. We note that the new young V09 PNe also generally have lower $r^{\prime}-\mbox{H}\alpha$ colours and higher $J-H$ and $H-K_s$ colours than the bulk of the normal PNe.

To conclude, we consider a higher probability for our candidates to be normal PNe if they are located high up in the IPHAS two-colour diagram and towards low $J-H$ and intermediate $H-K_s$ colours in the 2MASS two-colour diagram. However, for vyPNe we expect a lower position in the IPHAS two-colour diagram and larger $J-H$ and $H-K_s$ colour values.

\begin{figure}[!t]
   \centering
   \includegraphics[width=\columnwidth,angle=0]{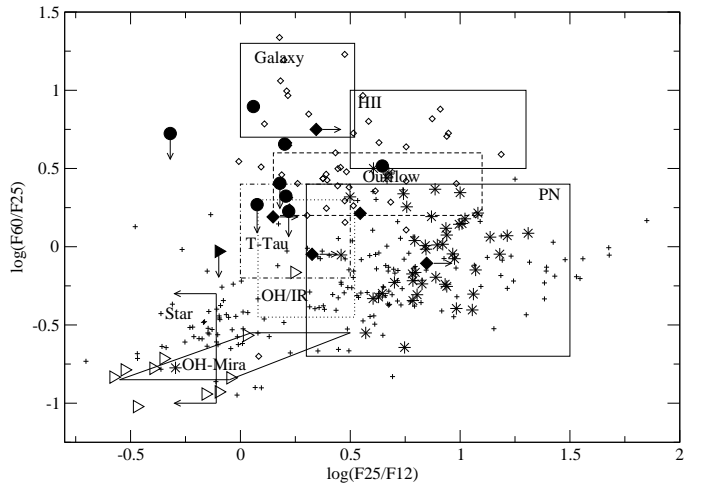}
      \caption{IRAS two-colour diagram. The symbols used are the same as in Fig.~\ref{fig:ip2mdiagram2}. Shown are also the locations of our candidates (open diamonds) and the \citet{belczynski00} symbiotic stars with an IRAS couterpart (open triangles).}
\label{fig:irasdiagram}
\end{figure}

\subsubsection{IRAS}

We cross-matched our candidate list with the IRAS catalogue, searching for possible matches in the IRAS Point Source Catalog, IRAS Faint Source Catalog, IRAS Cataloged Galaxies and Quasars, IRAS Serendipitous Survey Catalog and in the IRAS Small Scale Structure Catalog. When searching for IRAS data, a 16$\arcsec$ search radius was used. This choice is based on the positional accuracy of extended IRAS objects \citep{beichman88}. Of our 781 PN candidates, 51 are well measured (at least) in the $12\mu$m, $25\mu$m and $60\mu$m IRAS bands.

\citet{pottasch88} defined different areas marking the location of H\,{\sc ii} regions, OH/IR stars, PNe, normal late type stars, and galaxies in the IRAS $F(12\mu$m)$/F(25\mu$m) vs. $F(25\mu$m)$/F(60\mu$m) diagram, as shown in Fig.~\ref{fig:irasdiagram}. We also added to the diagram the locations of T-Tauri stars as studied by \citet{emerson87} and of outflow sources (both young and evolved stars with fast bipolar outflows) from \citet{zijlstra01}.

The locations of the newly discovered objects are again in line with expectations except for one IPHAS PN which is clearly above the PN area, in the Galaxy/H\,{\sc ii} region zone. However, the location of this object, IPHASXJ195248.8+255360, in the S2N2 diagram and its morphology (see Fig.~\ref{fig:ext_im}) clearly point to a PN nature. We consider a higher probability for our candidates to be a PN if they are located inside the PN box. But we also point out that a location outside of the PN box does not automatically rule out a PN nature.

\subsubsection{MSX + 2MASS infrared diagrams}

We also searched for MSX counterparts for our PN candidates. A 5$\arcsec$ search radius was used based on the MSX positional accuracy \citep{egan99}. 122 of our PN candidates have an MSX counterpart well measured at least in one of the MSX filters. 

\citet{lumsden02} studied various diagnostic diagrams combining MSX and 2MASS flux ratios in order to select mYSO candidates. For this purpose, they placed Herbig Ae/Be stars, mYSOs, methanol maser sources with radio emission (which trace massive star forming regions), compact H\,{\sc ii} regions, carbon stars, OH/IR stars, and PNe into the diagrams. We refer here to the diagrams shown in Figures 3, 4, 9 and 10 in \citet{lumsden02} as L1, L2, L3 and L4, respectively. In each of these diagrams we defined a line along the reddening vector which limits the zone where most of the known PNe in the diagrams are located. These cut lines in the L1, L2, L3 and L4 diagrams, respectively, are as follows:

\begin{displaymath}
\log[F(14\mu m)/F(8\mu m)] > -6.64\log[F(12\mu m)/F(14\mu m)]+2
\end{displaymath}

\begin{displaymath}
\log[F(14\mu m)/F(12\mu m)] < 0.56\log[F(21\mu m)/F(8\mu m)]-0.21
\end{displaymath}

\begin{displaymath}
\log[F(8\mu m)/F(K_s)] < 3.03\log[F(12\mu m)/F(8\mu m)]-0.86
\end{displaymath}

\begin{displaymath}
\log[F(K_s)/F(J)] < 29.29\log[F(21\mu m)/F(12\mu m)]-4.82.
\end{displaymath}

We studied the locations of our candidates, symbiotic stars,
  T-Tauri and post-AGB stars in the diagrams L1, L2, L3 and L4 and
  found that all of them separate PNe from T-Tauri and symbiotic stars
  quite nicely. However, we note that in all these diagrams, the
  PN zone overlaps with the zones occupied by compact H\,{\sc ii}
  regions and methanol maser sources with radio emission. There is
  also a partial overlap with the OH/IR star zone. In the L3 and L4
  diagrams, also some Herbig Ae/Be stars and mYSOs enter the PN
  zone. Nevertheless, as we expect T-Tauri and symbiotic stars to be
  the most frequent mimics in our catalogue, we consider these
  diagrams useful in weighting the probability that our candidates are
  a genuine PNe. However, the post-AGB stars are not separated from the rest of the objects in any of the diagrams. This can mean that the vyPNe would not be either.

\subsubsection{Discussion on the diagnostic diagrams}

We have studied in various diagnostic diagrams the locations of our candidates and the expected most frequent mimics, namely T-Tauri stars and symbiotic stars. The \citet{lumsden02} diagrams do not separate PNe from compact H\,{\sc ii} regions but the separation from other possible mimics is generally good, especially in the IRAS diagram and in the L1 and L2 diagrams. The IR data certainly can be used to further restrict the possibility that our candidates are genuine PNe in the cases where the necessary IRAS, MSX and 2MASS data for the candidate exist. 

We have also placed post-AGB stars in these diagrams as we expect that our candidate list includes, in addition to normal PNe, vyPNe or proto-PNe like those presented in V09. Based on the locations of the post-AGB stars as compared to PNe we expect that the vyPNe are found outside the most typical PN zones in IPHAS and 2MASS two-colour diagrams. The results in V09 support this observation. In these zones the confusion with mimics is higher and thus makes the discovery of vyPNe more difficult. In the IRAS and MSX diagrams the locations of post-AGB stars overlap with both PNe and normal late type stars. This might be an evolutionary trend, the most evolved post-AGB stars being located in the PN zone while the less evolved ones are located in the zone of late type stars. We find that we cannot separate the vyPNe from the normal late type stars based only on their infrared colours. Lastly, we note that in the IRAS diagram, the distribution of our candidates clearly differs 
 from the distribution of known PNe and post-AGB stars. Therefore, we conclude that most of our candidates with an IRAS counterpart are rather mimics than young PNe. We have included all available IR data in Table 1.

\section{Additional data}

\subsection{Radio}

Candidates with PN-like colours in the different optical/IR diagrams, showing an adequate morphology and associated with radio continuum sources, are almost certainly PNe. We searched for matches to our candidates by comparing the IPHAS and NVSS \citep[NRAO VLA Sky Survey,][]{condon98} radio images. 63 of the 781 IPHAS PN candidates were found to have an 1.4 GHz radio counterpart. We add the radio information to our candidate table.

\subsection{Diameters of the candidates}

Normal PNe are expected to be more extended than T-Tauri stars, symbiotic stars, post-AGB stars and other stars and we measured the diameters of all our candidates. For the 75 clearly extended candidates, the diameters and coordinates of the centre were measured in the IPHAS H$\alpha$ images using the FITS viewer ds9\footnote{http://hea-www.harvard.edu/RD/ds9}. For the 706 candidates that appear compact, we have measured their full width half maximums (FWHMs). For comparison, we measured the FWHMs of field stars in an area of 10$\arcmin$ surrounding the objects, selecting the stars within $\pm0.1$ magnitudes from our candidate, or as a minimum, 50 stars as close in magnitude to our object as possible. A 5\% trimmed and 2-$\sigma$ clipped mean of these defines the stellar FWHM and the RMS its error, as well as the error in the FWHM measurement of the object. We decided to use both trimming and sigma clipping for a robust measurement of the stellar FWHM. In order to claim that one candidate is resolved, we use a conservative limit of five sigma, i.e. we consider that the object is resolved if $FWHM(object)-FWHM(stars) > 5\times err\_FWHM(object)$.
Of the 706 compact objects, 124 turned out to be resolved using this criterion. Of these, 25 have a deconvolved diameter, $\sqrt{FWHM(object)^2-FWHM(stars)^2}$, smaller than 1 arcsec.

120 of the resolved candidates are
of size $\leq 5 \arcsec$, the size range for which our search method
is optimised. At diameters greater than $5\arcsec$, only a
  fraction of the nebulae where the semi-automated algorithm detected
  a small condensation is found. We hardly find any candidates with diameters smaller than $0.5\arcsec$, as expected for instrumental and observational reasons. There are 75 candidates with diameters in the range $0.5\arcsec - 2\arcsec$. This is the diameter range where young/compact/very distant PNe are expected, possibly accompanied by a small fraction of nebulae around T-Tauri stars and symbiotic stars. From $2\arcsec$ on, the number of objects clearly drops with increasing diameter. This can be attributed to the difficulty in detecting lower surface brightness objects at large distances due to interstellar extinction.


\begin{figure*}[!t]
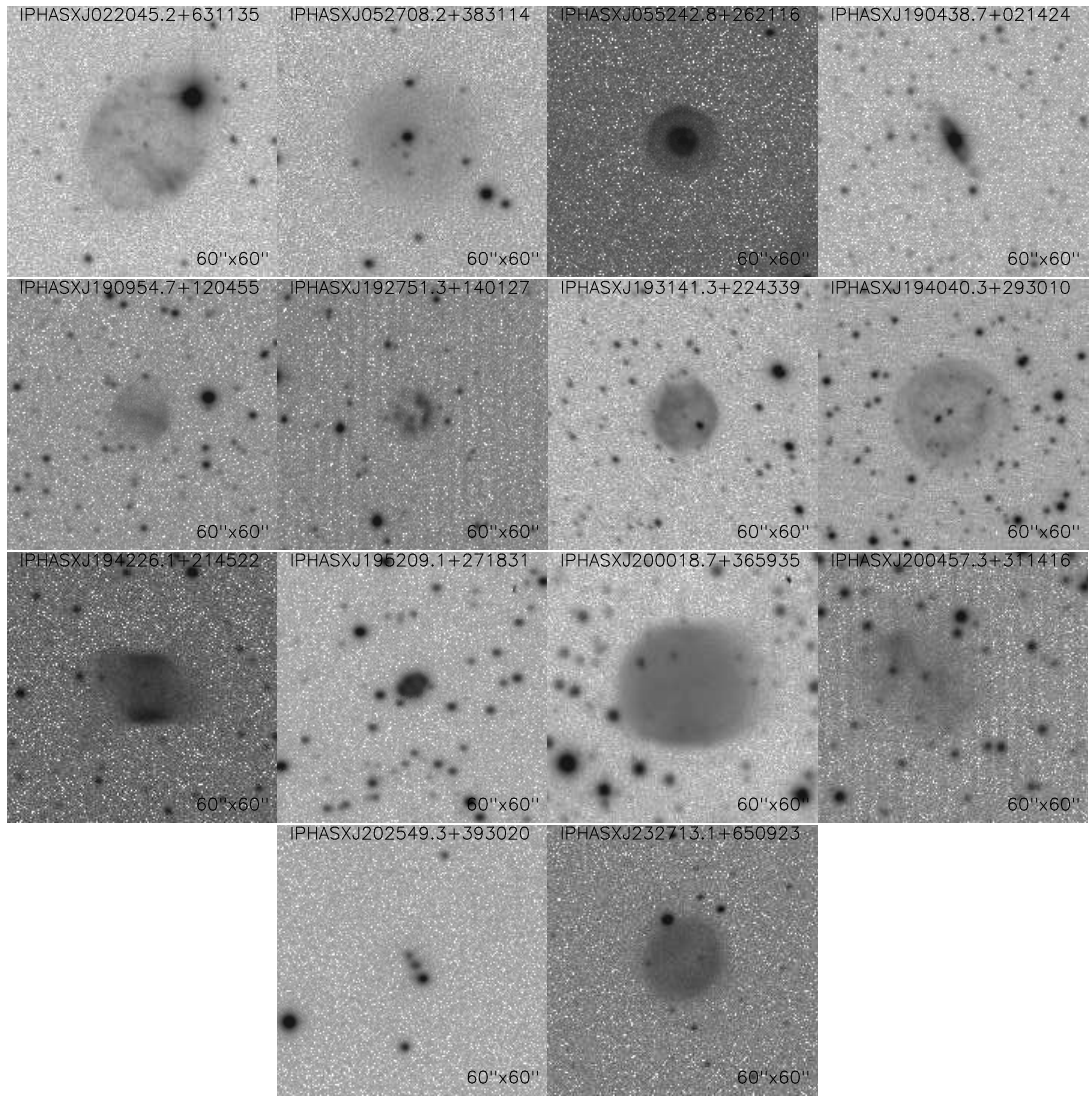

   \centering
\includegraphics[width=0.4\columnwidth,angle=-90]{Fig6a.eps}\includegraphics[width=0.4\columnwidth,angle=-90]{Fig6b.eps}\includegraphics[width=0.4\columnwidth,angle=-90]{Fig6c.eps}\includegraphics[width=0.4\columnwidth,angle=-90]{Fig6d.eps}
\includegraphics[width=0.4\columnwidth,angle=-90]{Fig6e.eps}\includegraphics[width=0.4\columnwidth,angle=-90]{Fig6f.eps}\includegraphics[width=0.4\columnwidth,angle=-90]{Fig6g.eps}\includegraphics[width=0.4\columnwidth,angle=-90]{Fig6h.eps}
\includegraphics[width=0.4\columnwidth,angle=-90]{Fig6i.eps}\includegraphics[width=0.4\columnwidth,angle=-90]{Fig6j.eps}\includegraphics[width=0.4\columnwidth,angle=-90]{Fig6k.eps}\includegraphics[width=0.4\columnwidth,angle=-90]{Fig6l.eps}
\includegraphics[width=0.4\columnwidth,angle=-90]{Fig6m.eps}\includegraphics[width=0.4\columnwidth,angle=-90]{Fig6n.eps}
\caption{IPHAS H$\alpha$ images of the 14 extended candidates with PN-like morphology and which have an entry in SIMBAD but whose nature is not confirmed.}
\label{fig:simext_im}
\end{figure*}

\begin{figure*}[!t]
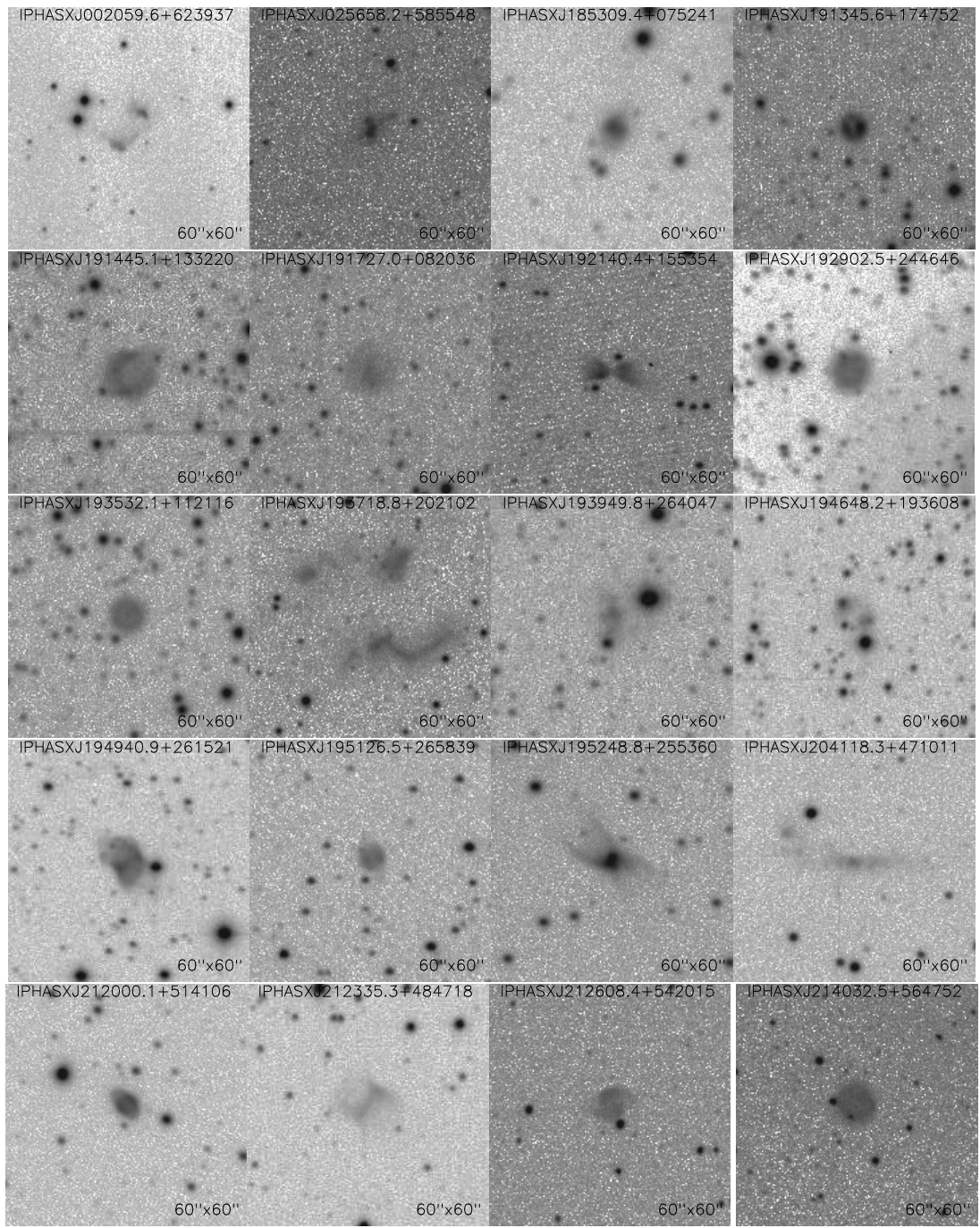

   \centering
   \includegraphics[width=0.4\columnwidth,angle=-90]{Fig7a.eps}\includegraphics[width=0.4\columnwidth,angle=-90]{Fig7b.eps}\includegraphics[width=0.4\columnwidth,angle=-90]{Fig7c.eps}\includegraphics[width=0.4\columnwidth,angle=-90]{Fig7d.eps}
\includegraphics[width=0.4\columnwidth,angle=-90]{Fig7e.eps}\includegraphics[width=0.4\columnwidth,angle=-90]{Fig7f.eps}\includegraphics[width=0.4\columnwidth,angle=-90]{Fig7g.eps}\includegraphics[width=0.4\columnwidth,angle=-90]{Fig7h.eps}
\includegraphics[width=0.4\columnwidth,angle=-90]{Fig7i.eps}\includegraphics[width=0.4\columnwidth,angle=-90]{Fig7j.eps}\includegraphics[width=0.4\columnwidth,angle=-90]{Fig7k.eps}\includegraphics[width=0.4\columnwidth,angle=-90]{Fig7l.eps}
\includegraphics[width=0.4\columnwidth,angle=-90]{Fig7m.eps}\includegraphics[width=0.4\columnwidth,angle=-90]{Fig7n.eps}\includegraphics[width=0.4\columnwidth,angle=-90]{Fig7o.eps}\includegraphics[width=0.4\columnwidth,angle=-90]{Fig7p.eps}
\includegraphics[width=0.4\columnwidth,angle=-90]{Fig7q.eps}\includegraphics[width=0.4\columnwidth,angle=-90]{Fig7r.eps}\includegraphics[width=0.4\columnwidth,angle=-90]{Fig7s.eps}
\includegraphics[width=0.4\columnwidth,angle=-90]{Fig7t.eps}
      \caption{IPHAS H$\alpha$ images of the 20 extended candidates with PN-like morphology.}
\label{fig:ext_im}
\end{figure*}

\section{Extended candidates} 

Although our search method is optimised to discover compact ($<$5$\arcsec$) PNe, we also found 81 extended objects with a diameter $>$5$\arcsec$. For these clearly extended objects their morphology can help in determining their nature. 36 nebulae show regular morphology so that from their appearance alone we can assume a higher probability that these objects are genuine PNe. Two of these were discovered by \citet{miszalski08}. We present here the images and short descriptions of the remaining 34 objects. Of these, 14 are known in SIMBAD but their classification is not well established and they deserve further study.

\subsection{Candidates classified in SIMBAD}

\noindent{\bf IPHASXJ022045.2+631135} Classified as a galaxy \citep{hau95}. However, it is hardly visible in the IPHAS continuum filters, which is an argument against the galaxy classification. Possibly an elliptical PN.

\noindent{\bf IPHASXJ052708.2+383114} Classified as a reflection nebula \citep{magakian03} but no detailed study of the object is available. Possibly a round PN.

\noindent{\bf IPHASXJ055242.8+262116} A CO source \citep{wouterloot93}. Possibly a round PN.

\noindent{\bf IPHASXJ190438.7+021424} Catalogued as an H$\alpha$-emission star \citep{kohoutek99}. \citet{condon99} have obtained an optical spectrum of this object and classify it as a PN. Preliminarily classified as a symbiotic star based on a MASH spectrum but more observations are needed. An elliptical PN or symbiotic star.

\noindent{\bf IPHASXJ190954.7+120455} Classified as a possible PN \citep{kronberger06} but no spectroscopic confirmation is available. Possibly a round or bipolar PN with a brightened waist.

\noindent{\bf IPHASXJ192751.3+140127} Classified as a possible PN \citep{preite88} but no spectroscopic confirmation is available. Possibly a round PN.

\noindent{\bf IPHASXJ193141.3+224339} Classifications of PN \citep{condon99}, Galaxy \citep{roman00} and possible PN \citet{preite88} but no spectroscopic confirmation is available. Hardly visible in the IPHAS continuum filters and thus unlikely to be a galaxy. Possibly a round/slightly elliptical PN.

\noindent{\bf IPHASXJ194040.3+293010} Classified as a possible PN by \citet{roman00} and as a PN by \citet{kronberger06} but no spectroscopic confirmation is available and hardly visible in the IPHAS continuum filters. Possibly a round PN.

\noindent{\bf IPHASXJ194226.1+214522} Classified as a possible PN \citep{kronberger06} but no spectroscopic confirmation is available. Possibly a bipolar/quadrupolar PN.

\noindent{\bf IPHASXJ195209.1+271831} Classified as PN (PN K 3-48). Many studies refer to this object as a PN but is listed as an H\,{\sc ii} region by \citet{acker94}. No spectra available in the literature. Possibly an elliptical PN.

\noindent{\bf IPHASXJ200018.7+365935} Classified as a possible PN \citep{acker94} but with no spectroscopic confirmation. Possibly a round PN.

\noindent{\bf IPHASXJ200457.3+311416} Listed as a 'Bright Nebula' \citep{kronberger06}. Possibly a round or a bipolar PN with brightened waist.

\noindent{\bf IPHASXJ202549.3+393020} Catalogued as an H$\alpha$ emission line star \citep{kohoutek97}, PN \citep{manchado89} and H\,{\sc ii} region \citep{garcia-lario97} but with no spectroscopic confirmation available. Possibly a bipolar PN.

\noindent{\bf IPHASXJ232713.1+650923} Classified as possible PN \citep{kronberger06} but with no spectroscopic confirmation available. Possibly a round PN.

\subsection{Candidates without any SIMBAD classification}

\noindent{\bf IPHASXJ002059.6+623937} A bipolar nebula.

\noindent{\bf IPHASXJ025658.2+585548} Possibly bipolar but a deeper image is needed to define the morphology.

\noindent{\bf IPHASXJ185309.4+075241} A roundish/slightly elliptical nebula.

\noindent{\bf IPHASXJ191345.6+174752} A round nebula. Brighter rims, especially the western one.

\noindent{\bf IPHASXJ191445.1+133220} Round nebula. Brightened rim.

\noindent{\bf IPHASXJ191727.0+082036} Faint nebula. Looks round but a deeper image is needed.

\noindent{\bf IPHASXJ192140.4+155354} Butterfly shaped nebula. Preliminarily classified as a symbiotic star based on a MASH spectrum but more spectra are needed to confirm the classification.

\noindent{\bf IPHASXJ192902.5+244646} A round nebula.

\noindent{\bf IPHASXJ193532.1+112116} A round nebula.

\noindent{\bf IPHASXJ193718.8+202102} A bipolar nebula. Only two irregularly shaped lobes visible.

\noindent{\bf IPHASXJ193949.8+264047} Shape unclear but if the bright central part is the waist of a bipolar nebula, the real extent can be larger than the 10$\arcsec$ listed in Table 1.

\noindent{\bf IPHASXJ194648.2+193608} A bipolar nebula. The two lobes are combined with an elliptical ring.

\noindent{\bf IPHASXJ194940.9+261521} A bipolar nebula. The waist is clearly brighter than the lobes.

\noindent{\bf IPHASXJ195126.5+265839} Faint round/elliptical nebula.

\noindent{\bf IPHASXJ195248.8+255360} Bipolar nebula. The waist is bright while the lobes are very faint.

\noindent{\bf IPHASXJ204118.3+471011} The morphology is not very clear but looks like a bright waist of a bipolar nebula with faint lobes only partially visible.

\noindent{\bf IPHASXJ212000.1+514106} An elliptical nebula. The NW and SE rims are brightened.

\noindent{\bf IPHASXJ212335.3+484718} A bipolar nebula, bright waist and only partly visible lobes.

\noindent{\bf IPHASXJ212608.4+542015} An 'S'-shaped nebula. Quadrupolar? Deeper imaging needed to define the morphology.

\noindent{\bf IPHASXJ214032.5+564752} A round nebula.


\section{The IPHAS PN candidate list}

We have collected all the information discussed above into a list of IPHAS PNe candidates (provided in electronic format only).  The table contains all the 1005 objects picked up by our selection algorithm. The columns of the table are as follows:

\noindent{\bf Name}: The IAU-registered sexagesimal, equatorial position-based source name in the form: IPHASJhhmmss.ss+ddmmss.s for stellar and IPHASXJhhmmss.s+ddmmss for extended objects. "J" indicates the position is J2000.

\noindent{\bf RA, DEC}: The right ascension and declination (J2000) of the object catalogued by IPHAS [deg].

\noindent{\bf Gal}: The galactic coordinates of the object in the form lll.l+bb.b where l and b stand for galactic longitude and latitude, respectively, truncated to one decimal place [deg].

\noindent{\bf r, r-i, r-Ha}: The IPHAS $r^\prime$ band magnitude, $r^\prime-i^\prime$ (if measured in $i^\prime$) and $r^\prime-\mbox{H}\alpha$ colours of the object. In the case of extended objects, the IPHAS magnitude/colour information corresponds to the brightest catalogued H$\alpha$ condensation of the object.

\noindent{\bf f\_r-Ha}: IPHAS photometry flag, meaning that the object is saturated in H$\alpha$.

\noindent{\bf cRA, cDE}: The right ascension and declination (J2000) of the measured centre for the object [deg]. This information is only presented for extended objects.

\noindent{\bf IRCat}: The near-IR catalogue (2MASS PSC, 2MASS XSC, or UKIDSS) from which the near-IR data of the object was taken.

\noindent{\bf J, J-H, H-Ks}: $J$ band magnitude, $J-H$ and $H-K_s$ colours of the object.

 \noindent{\bf IRASCat}: The IRAS catalogue (PSC/SSC/SSS) from which the IRAS data of the object was taken.  

\noindent{\bf f12, f25, f60, f100 }: IRAS fluxes at 12, 25, 60 and 100 $\mu$m [Jy].

\noindent{\bf fMSXa, fMSXc, fMSXd, fMSXe}: MSX fluxes at 8.28, 12.13, 14.65 and 21.34 $\mu$m [Jy].

\noindent{\bf fNVSS}: Radio flux at 1.4 GHz [mJy].

 \noindent{\bf sFWHM}: The stellar FWHM [arcsec]. 

 \noindent{\bf oFWHM}: The object FWHM [arcsec]. The FWHM information is provided only for the objects appearing quasi-stellar in IPHAS. 

 \noindent{\bf Diam}: Diameter for the clearly extended candidates [arcsec].

 \noindent{\bf PNIRAS}: In the PN zone in the IRAS diagram [yes/no].

 \noindent{\bf PNL1}: In the PN zone in the L1 diagram [yes/no].

 \noindent{\bf PNL2}: In the PN zone in the L2 diagram [yes/no].

 \noindent{\bf PNL3}: In the PN zone in the L3 diagram [yes/no].

 \noindent{\bf PNL4}: In the PN zone in the L4 diagram [yes/no].

 \noindent{\bf score}: The PN score as defined below.

\noindent {\bf f\_SIMBAD}: Flag indicating that we consider the SIMBAD type not well established.

 \noindent{\bf type}: The types listed in SIMBAD. 

\noindent{\bf SIMBAD}: The SIMBAD name.

\noindent{\bf crossref}: Cross reference with the other IPHAS and AAO-UKST H$\alpha$ Survey selected catalogues: C = IPHAS symbiotic star candidates \citep{corradi08}, S = IPHAS extended nebulae catalogue (S09), M = PN in MASH or MASH-II catalogues \citep{parker06,miszalski08}, W = IPHAS emitter catalogue \citep{witham08}.

\noindent {\bf class}: Preliminary classification based on follow-up spectroscopy. A detailed analyses of the objects and the spectra will be published elsewhere. Objects with a MASH-II classification have been followed up by the MASH project. (IPHAS = spectrum taken but not yet studied, IPHAS PN/MASH-II PN = likely planetary nebula, IPHAS nebula = possible PN, IPHAS/MASH-II star = H$\alpha$ emission line star, IPHAS/MASH-II symbiotic = likely symbiotic star, See S09 = will be published in S09, See V09 = was studied in V09).

In addition the magnitude and colour errors are given in the columns {\bf e\_x}\footnote{For the near-IR data an error value of zero indicates that the magnitude, or one of the magnitudes defining the colour, is a 95\% confidence upper limit \citep{cutri03}.} and the flux qualities in the columns {\bf q\_x}\footnote{For IRAS PSC and SSC: 3=high quality, 2=moderate quality, 1=upper limit. For IRAS SSS: A=high quality, B=intermediate quality, F=low quality \citep{beichman88}. For MSX: 4 = excellent, 3 = good, 2 = fair, 1 = limit \citep{egan99}.} for the x data.

The PN score is defined from the source location in the IPHAS, 2MASS and other IR diagnostic diagrams in the following way: In the IPHAS two-colour diagram, the distance in magnitudes to the lowest selection line is calculated. To this we add the calculated distance in magnitudes from the selection line in the 2MASS two-colour diagram. If the object is located to the left of this selection line the distance is simply set to zero. If in addition the object is located in the PN zone in any of the infrared diagrams, its score is increased by one for each IR diagrams where the object is in the right zone. So, in summary, the higher the PN score of the object is, the higher we consider the possibility that this candidate is really a PN. For example, for the known PNe in the candidate list, the average PN score is 3.6 while for the other types of objects classified in SIMBAD the average is 2.5. For the new IPHAS PNe the average score is 3.3 and for the new IPHAS emission line
  stars 1.9.

However, the PN score alone cannot obviously be used to discard the possibility that a candidate is a PN. For example, the PN score of the known, extended PN, PN G036.9-02, is only 0.52. Its H$\alpha$ excess is low due to an underlying star which was picked up by our selection algorithm and it is not an IRAS nor MSX source which explain the low score.

In addition to the PN score, the radio and diameter data give additional constraints on the probable nature of the object.

\section{Summary and discussion}

IPHAS provides a powerful database for PN searches in the Northern Galactic Plane. Thanks to its high resolution and depth, especially the two extremes of the PN evolution, young, compact PNe and old, faint, and extended PNe are explored. The visual search, optimal for the discovery of very extended PNe, is underway (S09) while here we have presented the semi-automated search for PNe based on the IPHAS photometric catalogue. This method is optimised to discover small-sized PNe but can also discover some extended PNe. In addition to normal PNe, we expect that our candidate list includes vyPNe and proto-PNe.

We present a list of 1005 objects selected from the initial catalogue of $\sim600$ million IPHAS detections. The SIMBAD database and literature study reveals that 224 of these are actually known objects, 149 being confirmed PNe. This leaves us with 781 PN candidates. About 25\% of the selected candidates are slightly extended and the clearly extended ones ($>5\arcsec$) overlap with the visual search by S09 for extended PNe. This ensures that the combined IPHAS PN search is as complete as possible.

When selecting the PN candidates, we have aimed to be complete rather than having a clean selection of candidates. The first results from our follow-up spectroscopy reveal that 38 out of 69 candidates observed show nebular lines, whereas the remaining candidates are H$\alpha$ emission line stars, most probably YSOs. 
In order to help in the selection of good PNe candidates we have studied the locations of the candidates in a number of IR diagnostic diagrams. Further, we provide additional information like the sizes of the objects and the available radio data for them.

The probability that a candidate is a genuine, normal PN is higher if it is located in the PN area in the IPHAS and IR diagrams and if it is extended in H$\alpha$. All this information is provided in our candidate list and the colour-colour diagram locations are used to define a PN score whose value is higher for candidates with properties like the known, normal PNe.

However we would like to point out that exploring the objects not fulfilling the ''normal PN'' criteria listed above can be surprisingly productive in discovering objects in rare PN phases, as is shown by the very young PNe discovered and discussed in V09.


\begin{acknowledgements}

K.~V., A.~M., and R.~L.~M.~C. acknowledge funding from the Spanish AYA2007-66804 grant. In addition, K.~V. acknowledges the grant from the Magnus Ehrnrooth foundation, Finland. M.~R. and G.~D.-I. acknowledge support from Mexican CONACYT project 50359-F. This paper makes use of data obtained as part of IPHAS carried out at the INT as well as INT, WHT, NOT and SPM2.1m spectroscopic data. The INT and WHT are operated by the Isaac Newton Group and NOT by NOTSA on the island of La Palma in the Spanish Observatorio del Roque de los Muchachos of the Instituto de Astrof\'{i}sica de Canarias. SPM 2.1m is operated by UNAM at the OAN of San Pedro M\'artir Observatory, Mexico. All IPHAS data are processed by the Cambridge Astronomical Survey Unit, at the Institute of Astronomy in Cambridge. We also acknowledge use of data products from the 2MASS, which is a joint project of the University of Massachusetts and the Infrared Processing and Analysis Centre/California Institute of Technology 
 (funded by the USA's National Aeronautics and Space Administration and National Science Foundation).This research made use of data products from the Midcourse Space Experiment.  Processing of the data was funded by the Ballistic Missile Defense Organization with additional support from NASA Office of Space Science.  This research has also made use of the NASA/ IPAC Infrared Science Archive, which is operated by the Jet Propulsion Laboratory, California Institute of Technology, under contract with the National Aeronautics and Space Administration.

\end{acknowledgements}

\bibliographystyle{aa}
\bibliography{oma}

\end{document}